\documentclass[titlepage,twoside]{article}
\usepackage[american]{babel}
\usepackage[utf8]{inputenc}
\usepackage[T1]{fontenc}
\usepackage[backend=biber,useprefix=true,style=apa,url=true,dashed=true,doi=false,eprint=false]{biblatex}

\usepackage{csquotes}

\DeclareLanguageMapping{american}{american-apa}

\DeclareUnicodeCharacter{0301}{\'{e}} 

\usepackage[breaklinks, colorlinks=true, allcolors=black]{hyperref}
\usepackage{amsmath}
\usepackage[rightcaption]{sidecap}
\usepackage{caption}
\usepackage{sidecap}
\usepackage{placeins}
\usepackage{float}
\usepackage{subfigure}
\usepackage{MnSymbol} 
\usepackage{enumerate} 
\usepackage[table]{xcolor} 
\usepackage{multirow} 
\usepackage{adjustbox} 
\usepackage{graphicx} 
\usepackage{longtable} 
\usepackage{booktabs} 

\usepackage{ragged2e}

\usepackage{verbatim}

\graphicspath{ {./figures/} }

\usepackage{textcomp}
\usepackage{gensymb}

\usepackage{indentfirst}

\makeatletter
\renewcommand*{\@fnsymbol}[1]{\ensuremath{\ifcase#1\or *\or \dagger\or \ddagger\or
   \mathsection\or \mathparagraph\or \|\or **\or \dagger\dagger
   \or \ddagger\ddagger \else\@ctrerr\fi}}
\makeatother

\usepackage{fancyhdr,lipsum}
\begin{document}





\title{VR-NRP: A Virtual Reality Simulation for Training in the Neonatal Resuscitation Program}


\author{Mustafa Yalin Aydin$^1$ \and Vernon Curran$^2$ \and Susan White$^3$\and Lourdes Pe\~na-Castillo$^1{^,{^4}}$\and Oscar Meruvia-Pastor$^1{^,}$\thanks{Corresponding author: Oscar *at* MUN *dot* ca}}

\date{
	$^1$Department of Computer Science, Faculty of Science, Memorial University of Newfoundland, Canada \\ 
	$^2$Faculty of Medicine,Memorial University of Newfoundland, Canada\\ 
    $^3$Perinatal Program NL (PPNL), Newfoundland \& Labrador Health Services, Canada\\ 
    $^4$Department of Biology, Faculty of Science, Memorial University of Newfoundland, Canada%
}

\maketitle

\abstract{
The use of Virtual Reality (VR) technologies has been extensively researched in surgical and anatomical education. VR provides a lifelike and interactive environment where healthcare providers can practice and refresh their skills in a safe environment. VR has been shown to be as effective as traditional medical education teaching methods, with the potential to provide more cost-effective and convenient means of curriculum delivery, especially in rural and remote areas or in environments with limited access to hands-on training. In this sense, VR offers the potential to be used to support resuscitation training for healthcare providers such as the Neonatal Resuscitation Program (NRP). The NRP program is an evidence-based and standardized approach for training healthcare providers on the resuscitation of the newborn. In this article, we describe a VR simulation environment that was designed and developed to refresh the skills of NRP providers. To validate this platform, we compared the VR-NRP simulation with exposure to 360-degree immersive video.  We found that both VR technologies were positively viewed by healthcare professionals and performed very similarly to each other. However, the VR simulation provided a significantly increased feeling of presence. Furthermore, participants found the VR simulation more useful, leading to improved experiential learning outcomes. Also, participants using VR simulation reported higher confidence in certain NRP skills, such as proper mask placement and newborn response evaluation. This research represents a step forward in understanding how VR and related extended reality (XR) technologies can be applied for effective, immersive medical education, with potential benefits for remote and rural healthcare providers.
\noindent\textbf{Keywords:} Virtual Reality (VR), Neonatal Resuscitation Program (NRP), VR-NRP, Medical Training, Medical Education, VR Simulation
}

\section{Introduction}

Virtual Reality (VR) is a new and innovative technology being increasingly used for medical education and training. The use of VR technologies has been extensively researched in surgical and anatomical education with positive results ~(\cite{Jiang:2022:VRMedEdScopeRev,Makinen2020,Curran:2023:IntegrativeRev}). This is because VR provides a lifelike,  interactive and simulated environment where healthcare providers can practice and refresh their skills. The term VR was first coined in 1987 by Jaron Lanier. VR is defined as “a computer-generated, interactive, three-dimensional environment in which a person is immersed”~(\cite{Lele2011}). In VR, computer graphics systems are used in combination with various display and interface devices to provide the effect of immersion in an interactive 3D computer-generated environment, called a virtual environment (VE)~(\cite{Pan2006}). In a virtual environment, users can navigate, move around, and explore the features of a 3D scene. Users are also able to interact with the environment, meaning they can select and manipulate objects in the scene, receive inputs, and produce outputs into the system. For example, users are able to walk through a forest (navigate) and grab and examine a flower found in the forest (interact)~(\cite{Gutirrez2008}). In order to interact with the virtual environment, tools such as Head-Mounted Displays (HMDs), VR controllers, 3D glasses, gloves, joysticks and other devices are used~(\cite{Alhalabi2016, Rengganis2018}). In recent years, HMDs have become the primary method to experience VR, as they have become more affordable and can provide users with an immersive experience with realistic images and sounds that replicate the actual physical environment~(\cite{Farshid2018}). Most current HMDs do not allow users to see the outside view of the real world, but rather, users see the virtual environment instead~(\cite{Ghobadi2020}). VR systems are now used in many different fields, such as for entertainment, architectural design, education, learning and social skills training, simulations of surgical procedures, assistance to the elderly, and psychological treatments~(\cite{Cipresso2018}). HMDs have also been tested in training for sports, and particularly in industries where new employees are able to receive “risk-free” training using VR systems. Since the development of VR, many VEs have been created and applied to educational tasks in knowledge areas such as mathematics, language, business, health, computer science, and project management~(\cite{Checa2019}).

VR, Augmented Reality (AR), and Mixed Reality (MR), all encompassed under the umbrella term Extended Reality (XR), have been proposed as suitable technologies to facilitate learning. One of the reasons for this is that virtual learning environments (VLE) provide rich teaching patterns and instructional content. XR also helps to improve learners’ ability to analyze problems and explore new concepts, when properly designed. By being immersive, interactive, and imaginational, VR can provide virtual learning spaces that can be accessed by all kinds of learners~(\cite{Pan2006}).

VLEs can enhance education in at least three ways: 1) by presenting multiple perspectives, 2) by providing situated learning, and 3) by supporting skills transfer. Studies have shown that VLEs improve students’ performance significantly. A study by~(\cite{Alhalabi2016}), tested engineering students’ performance using three major VR systems and found that, compared to the controlled non-VR group, students performed significantly better when using any VR system than no VR. Additionally, the HMD VR system was found to have superior results than other VR systems. This study showed that the more students are immersed in the environment, the more they learn and the better they perform.

Research has also shown that knowledge is better retained when the learning material is delivered in a stimulating manner, as it is usually done in VLEs~(\cite{Lewis1994, Yardley2012}). One of the ways VR is used in learning in an engaging manner is through “serious games”. Serious games are entertaining activities from which users can also learn and be educated/trained in well defined areas and tasks~(\cite{Checa2019}). One of the benefits of serious games is that situations which could not otherwise be done in real life, such as ethical dilemmas and/or dangers, can be recreated in serious games, for training purposes. Virtual Reality Serious Games (VRSGs) provide the added benefit of an immersive VR environment, which improves user experiences and, therefore, knowledge acquisition. VRSGs have also been explored in industry and sports for skills training. Other areas where VRSGs have been used include training at educational institutions, such as sensitivity to bullying and motivating presentations for teachers and in the medical sector, especially skills improvement and knowledge acquisition developed for hospital staff~(\cite{Checa2019}).

According to recent research, education which incorporates VR technologies is perceived by learners to be more engaging and enjoyable, with higher levels of satisfaction and perceived usefulness~(\cite{Moro2017, Pantelidis2009}). Checa and Bustillo’s review of VRSGs for training and education showed that interactive experience was preferred~(\cite{Checa2019}). The review also concluded that user satisfaction with VRSG experience was higher than any other learning methodologies. Overall, VRSGs provide an effective learning platform with high user satisfaction and have a vast potential for application in many different areas, due to its affordability, technological development, providing a presence feeling, and  versatility.

\section{Motivation}

According to the World Health Organization (WHO), in 2016, an estimated 2.6 million newborns died in the first 28 days of life – 1 million of them died on the first day~(\cite{WHO2020}). One of the leading causes of neonatal death is birth asphyxia. Up to 10\% of newborns need help to begin breathing at birth with approximately 1\% requiring extensive resuscitation measures to restore cardio-respiratory function. In this context, the ability of healthcare providers to perform high-quality newborn resuscitation is essential for decreasing neonatal mortality~(\cite{WHO2020}).  The need for such interventions is relatively rare but highly stressful, and even for experienced healthcare professionals, medical errors or deviations from the resuscitation procedures may occur~(\cite{Ghoman2019}). Therefore, neonatal resuscitation training is crucial for enhancing the quality of care and positive patient outcomes~(\cite{Lee2011}). To address this, booster and refresher sessions involving ``mock codes'' have been shown to improve knowledge and skill updating~(\cite{Bender2014}). However, access to simulation equipment and organization of timely sessions may be challenging for health professionals in smaller facilities or rural locations.

The Neonatal Resuscitation Program (NRP) is a standardized, evidence-based approach for training health care providers on the resuscitation of a newborn. It was developed in 1987 by the American Heart Association (AHA) and the American Academy of Pediatrics (AAP), to identify infants at risk of respiratory depression and provide high-quality resuscitation~(\cite{Raghuveer2011}). We identify the NRP program to be an ideal area for testing whether VR can be used for refreshing and training practitioners in the context of continuous education, as it has been shown that VR provides an enhanced sense of presence and immersion amongst learners in clinical sciences training~(\cite{Jensen2017}). To maintain current Provider status, the Canadian Paediatric Society requires practitioners to successfully complete an in-person NRP Provider course at a minimum of every 2 years~(\cite{CPS:NRP}). VR training could be used as a complementary tool to refresh the skills of NRP providers during the 2 year term between Provider courses. 



\section{Related Work}
\subsection{Virtual Reality Usage in Medical Education and Training}
One of the major uses of VR is in the medical field, with research showing that, by 2018, about one in every five of all  VR publications were in surgery, rehabilitation and clinical neurology categories~(\cite{Cipresso2018}). 
VR is widely used in research on new ways of applying psychological treatment or training, such as for treatments and management of phobias~(\cite{Cipresso2018}). VR is also being utilized in medical education. This is largely because it provides opportunities for users to learn without adverse real-world consequences. For example, medical students can be trained to perform various surgical skills using surgical simulations via robotics and VR systems~(\cite{Berntsen2016}). This removes the pressure of learning on real patients while simultaneously providing training for various possible surgical scenarios.

According to a review by~(\cite{Makinen2020}), three types of VR technologies are being used for healthcare learning: haptic simulators, computer-based simulations, and HMD simulations. Of the three, haptic simulators were found to be used the most for healthcare learning, followed by computer-based simulators as the second most used technologies. Haptic simulators are mostly used in surgical training or medical education, as well as in nursing studies and dental studies~(\cite{Makinen2020}). The review also showed that different haptic or surgical simulators are used in training for surgery, lumbar punctures~(\cite{Farber2008}), intravenous (IV) insertions~(\cite{Reznek2002}), mandibular fracture reductions~(\cite{Girod2016}), and urethral catheterizations~(\cite{Johannesson2012}).

In contrast, computer-based simulators were mostly used in nursing education or in training healthcare trainees such as residents. The most commonly used computer platform for learning was the Second Life online virtual world. Computer-based serious games such as Virtual Emergency TeleMedicine (VETM), Body Interact and Home Healthcare Virtual Simulation Training Systems were also used for training purposes~(\cite{Nicolaidou2015, Padilha2018, Polivka2019, Verkuyl2019}). Lastly, HMDs including HTC Vive, Microsoft HoloLens, and Meta Quest, are also used for training healthcare faculty and nursing students  Some medical areas where HMDs have been used include imaging training~(\cite{Suncksen2018}), teaching surgical procedures~(\cite{Bracq2019}), training on urinary catheterizations~(\cite{Butt2018, KardongEdgren2019}) and teaching about medication withdrawal~(\cite{Vottero2014}). 

Simulation usage in neonatal resuscitation has been increasing~(\cite{Garvey2020}), as simulations help healthcare personnel maintain and refresh current knowledge.  In addition, studies conducted previously have shown that resuscitation training through simulations is more effective and suitable for enhancing teamwork skills among trainees~(\cite{Langhan2009}).

In terms of virtual reality simulations, there are VR serious games about NRP training aimed at improving participants’ learning and their abilities. One is called eHBB, developed by the University of Washington, USA and Oxford University, UK~(\cite{eHealthForEveryone}). The VR game can be accessible by mobile smartphones and also low-cost VR devices such as Google cardboard. It provides a non-interactive virtual environment and lets users observe the procedures of neonatal resuscitation through their smartphones. Its purpose is to provide an opportunity for continuous learning for new skill development, as well as to help maintain training fidelity. Overall, eHBB was reported to be easy to use, educational, and to enable learning without real-life stress~(\cite{Ghoman2019}).

Another example of a VRSG aimed at improving neonatal resuscitation is Life-saving Instructions For Emergencies (LIFE)~(\cite{oxLife2018}), developed by Oxford University, UK. LIFE is both a mobile and VR game that enables users to accomplish the procedure while using their own smartphones, thereby allowing healthcare workers to experience the game even within low-cost settings~(\cite{oxLife2020}). The game included a section regarding neonatal resuscitation where users must match the equipment correctly in a virtual rural hospital~(\cite{Ghoman2019}).

The Compromised Neonate Program is another immersive VR application developed by the University of Newcastle, Australia, for teaching neonatal resuscitation skills to midwifery students~(\cite{UniversityNewcastle2017}). In this VR application, learners must complete neonatal resuscitation tasks successfully in the virtual hospital environment~(\cite{Jones:2022,Fealy:2023}). 


VRSGs can be divided into two categories: information-based (non-interactive) immersive VR applications, and immersive interactive VR applications using hand held controllers. According to various researchers, information-based VRSGs about NRP are not very effective~(\cite{Yeo2020}), while interactive VR games seem promising, but little research has been done to evaluate their effectiveness~(\cite{Ghoman2019}). The work we present in this article aims at filling this gap in research by developing a new VR neonatal resuscitation application and evaluating potential NRP skill improvement by means of a controlled study.

\cite{Curran2021} explored the use of 360$^{\circ}$ NRP training videos using VR HMDs for neonatal resuscitation training  to explore healthcare providers’ experiences and perceptions. Overall, a high level of acceptance of the technology was reported by healthcare providers. In addition, it was found that 360$^{\circ}$ video provided an enhanced immersion experience in resuscitation scenarios, a strong sense of presence, and produced a greater level of interest. The main educational benefits reported included its usefulness for self-learning and as a supplement of traditional teaching methods and resources. In our case, we used 360$^{\circ}$ NRP training videos to compare against a computer-generated VR simulation of an NRP training session. 

In this article we aim to:
\begin{itemize}
    \item Explore the feasibility and implications of using computer-generated immersive VR simulations delivered through head-mounted displays (HMDs) to support neonatal resuscitation booster or refresher learning. 
    \item Explore  learners’ perspectives on using VR HMDs as a learning modality for neonatal resuscitation training.
    \item Describe in detail the proposed VR simulation for NRP training to facilitate scientific reproducibility and exchange.
    \item Examine the effect of VR HMDs usage on learning outcomes in neonatal resuscitation.
    \item Explore barriers and enablers towards the integration of VR HMDs as instructional or self-learning resources for healthcare providers trained in NRP. 
    \item Describe potential avenues for future research in VR-NRP simulation.    
\end{itemize}

\section{Methodology}
\label{section:Methodology}

In the sections below we describe the prototype developed and our design choices. 

\subsection{VR Simulation}

For this project, we used the Unity 3D game engine (version 2020.02f)~(\cite{Unity}) and its development environment for the visual design of the VR simulation prototype. An HTC VIVE Pro~(\cite{VIVEHeadsets}) which runs with SteamVR was used as the HMD platform. The VR simulation allows users to interact with the VR environment and move around, so we expect VR to provide a stronger sense of presence  than than experienced with the 360$^{\circ}$ videos.

\subsubsection{VR environment}
\begin{figure}[ht]
\centerline{\includegraphics[width=1\textwidth]{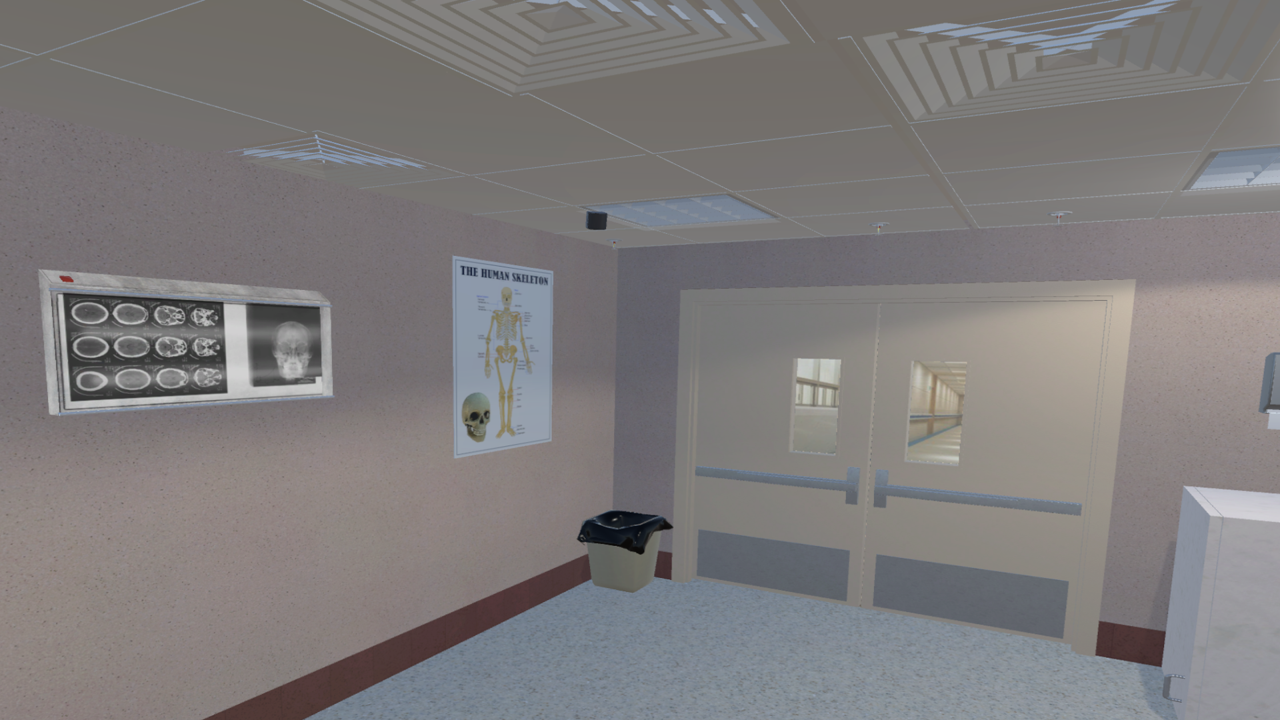}}

\caption{Sample view of the NICU welcome room. The ambient sounds are typical of the interior of a medical facility.}
\label{figure:hallways}
\end{figure}

The higher the realism in training simulations, the more effective the simulation is in terms of refreshing the experience from a trainee’s perspective~(\cite{Howard:2024,Choi:2017}). For this reason, we created a virtual environment to simulate the real working environment of NRP trainees (also referred to as learners or users). 
We drew our inspiration from the Eastern Health Neonatal Intensive Care Unit (NICU). The walls, corridors, floors and ceiling of the space in the simulation were designed to resemble the actual environment and we added relevant devices, posters, and furniture so that users could feel a strong sense of presence in a NICU (see Figure~\ref{figure:hallways}).   

\subsubsection{Spatial Arrangement}
In this simulation, there are three rooms in total. When the program starts, the user's first room is the welcome room, followed by the tutorial room and then the simulation room. This also reflects the expected sequence in which users are meant to move across rooms. All rooms are similar to each other but with different equipment and objects. For example, while there are no objects to interact with in the welcome room, the infant and doctor avatars in the tutorial room are the same as the infant and doctor avatars in the simulation room.

\subsubsection{First Room (Welcome Room)}
\begin{figure}[tbh]
\centerline{\includegraphics[width=1\textwidth]{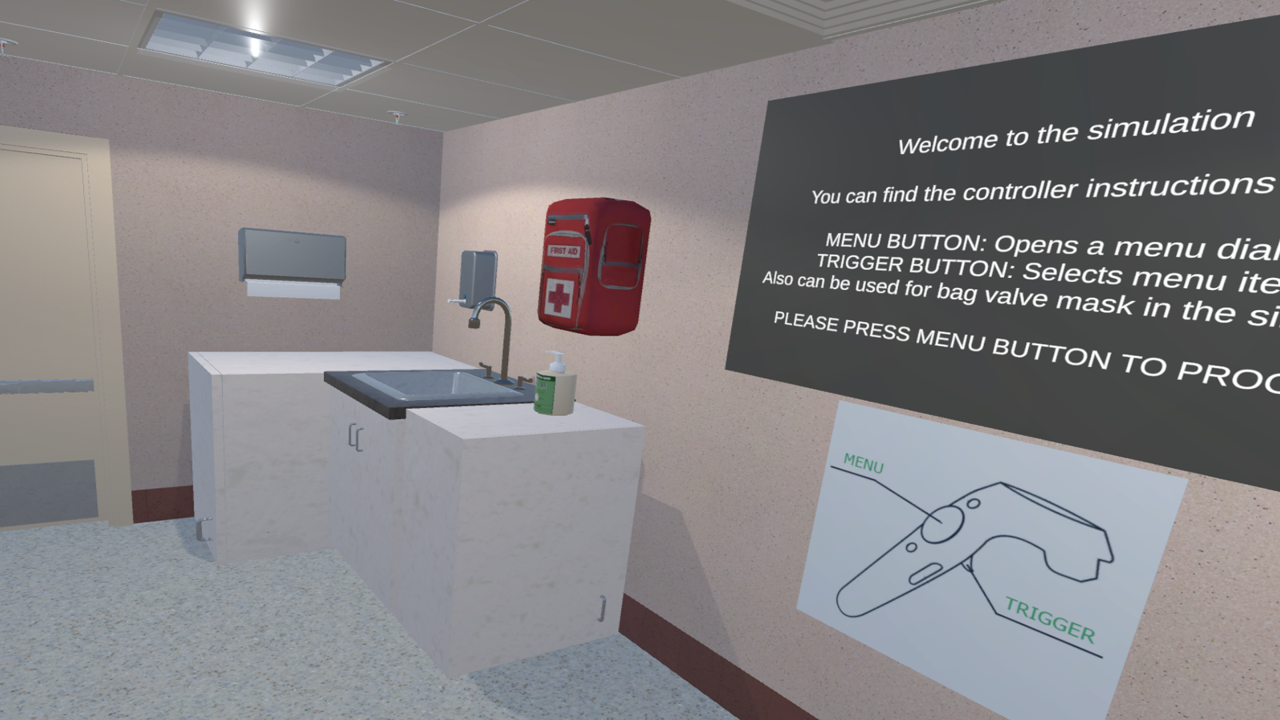}}

\caption{Instructions on how to use the 
controllers are shown in the welcome room, along with a brief audio message.}
\label{figure:simulation_tutorial_room}
\end{figure}

The welcome room is the room where the user gains familiarity with the simulation environment, has the opportunity to learn the controls, and checks that all equipment is working. When the program starts, it automatically starts from this room, and there is no object that the user can interact with because the purpose is to introduce the user to the environment. On the wall in front of the user, there is an information canvas where the functions of the buttons on the equipment and the controller are explained (see Figure~\ref{figure:simulation_tutorial_room}).


The first room is also the room where the sound, which is of great importance in simulations, is controlled. In order to reflect the hospital ambience, background “hospital sound” is added and set at a low level such that it is not disturbing/distracting the user. It is important for the user to hear the sounds from the simulation clearly because most of the information and guidance in the simulation is done with sound effects. For example, it is important for the users to hear what the doctor avatar is saying as he is guiding the user. Another example includes the sound when the bag valve mask is squeezed while positive pressure ventilation (PPV) is applied, which gives feedback that the bag is squeezed appropriately and also gives an immersive experience. Each user is asked if they hear the sounds clearly before they start the simulation.

\subsubsection{Second Room (Tutorial Room)}

\begin{figure}[ht]
\centerline{\includegraphics[width=1\textwidth]{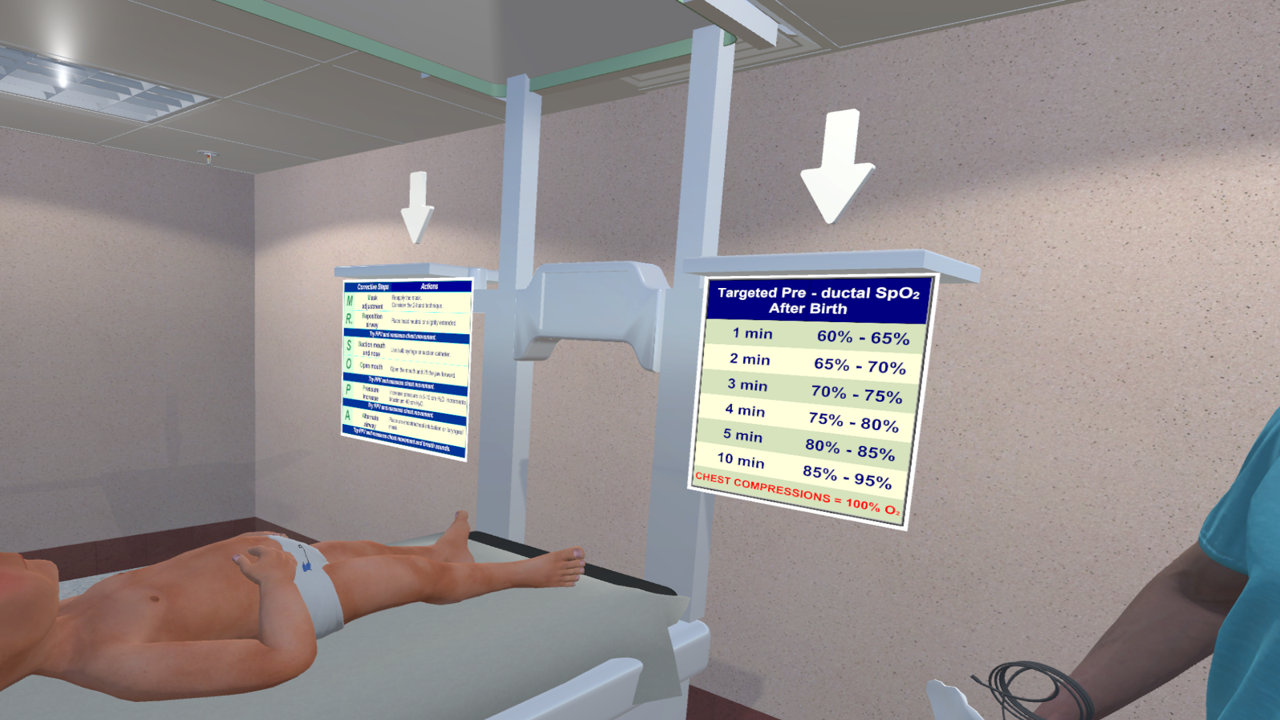}}
\caption{MR.SOPA and targeted pre-ductal saturation posters were hung by the sides of the infant warmer in the tutorial room.}
\label{figure:posters}
\end{figure}

The second room is called the “Tutorial Room”. The structure of this room is the same as the Welcome Room. However, unlike the welcome room, there is an infant warmer in front of the user, an infant on the infant warmer, a doctor's avatar who welcomes the user, and NRP information posters on the infant warmer that the user can refer to during the simulation (Figure~\ref{figure:posters}). To draw the attention of the user to these posters, arrows pointing to the posters have been placed on top of them. The doctor avatar is located on the user’s right side, next to the infant warmer. The doctor avatar will also help the user perform PPV during the neonatal resuscitation procedure as part of a team-approach to NRP. The user can move the bag valve mask with the help of the HTC VIVE Tracker in their left hand. The controller in the user's right hand represents the user's hand, as in other rooms. The doctor avatar looks at the user during the conversations for an immersive experience. This feature has been implemented with Unity's “Animation Rigging” package. The head of the doctor avatar is set to follow the VR camera, that is, the user.

\subsubsection{Third Room (Simulation Room)}
The third room is called the “Simulation Room”. After the user understands the system in the welcome room and masters the equipment in the tutorial room, they teleport to this room to start the neonatal resuscitation training simulation. The simulation starts shortly after they are teleported to this room. The simulation room design is the same as the other rooms. The placement of the objects is the same so that the user feels familiar. Unlike other rooms, there is a monitor to display heart rate and oxygen saturation on the left side of the user, where they can monitor the infant's vital signs. During the training simulation, the data received from the pulse oximeter attached to the infant's right wrist and ECG leads is reflected on this monitor. When the user turns their head to the left, they can easily observe this data.

Unlike the other rooms, there are two different objects on the avatar of the doctor. The first is the stethoscope that he will use when necessary during the simulation and the pulse oximeter that he puts on the infant. The visibility of these objects is initially set as invisible, and  then, when necessary in the later phases of the training, set as visible.

Another object that only exists in the simulation room is the analog clock on the wall opposite the user. This clock is a real-time clock and has been set for the user to follow the neonatal resuscitation process and to check the time when necessary.

\begin{figure}[ht]
\centerline{\includegraphics[width=1\textwidth]{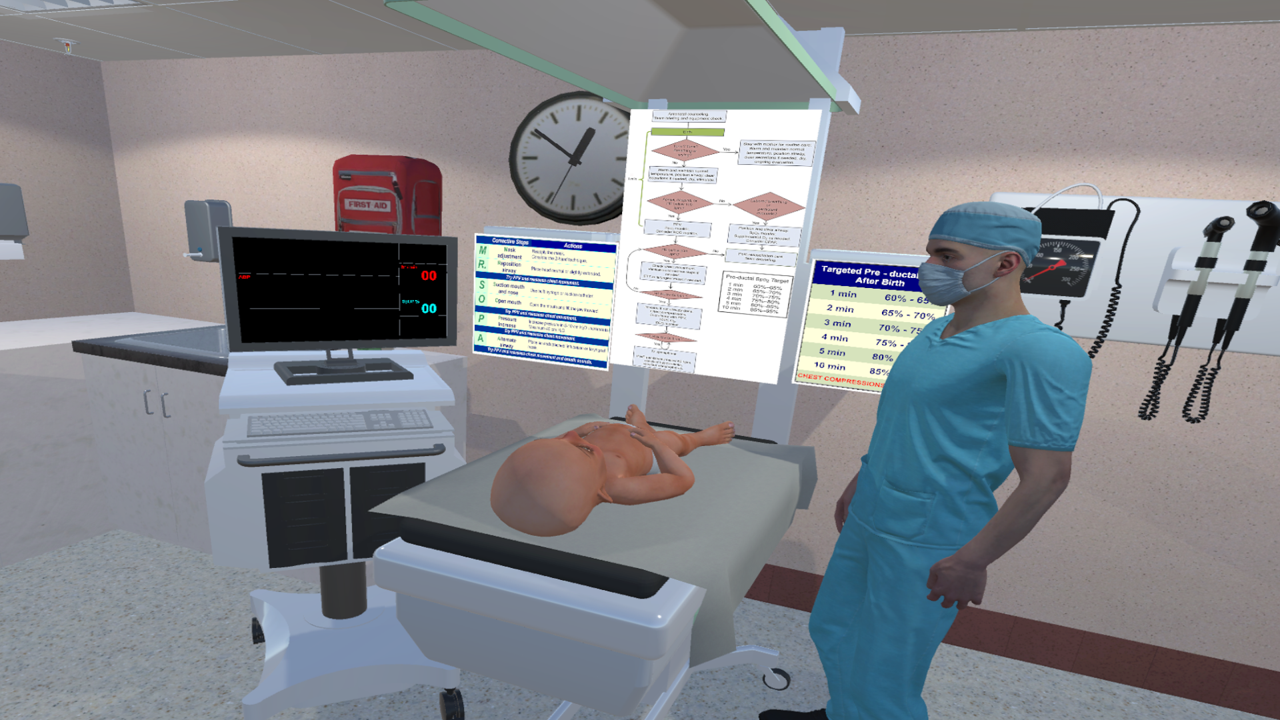}}
\caption{A sample view of the simulation room from the user's perspective.}
\label{figure:simulation_general}
\end{figure}

NRP (7th edition) posters were hung on the infant warmer to assist healthcare workers in the neonatal resuscitation procedure (Figure \ref{figure:simulation_general}). In addition, a poster of “mask adjustment, reposition the airway, suction mouth/nose, open the mouth, increase the peak inspiratory pressure (PIP), altenate airway" (MRSOPA), which is an important neonatal resuscitation action sequence, is placed on the front of the infant warmer, close by the user. Finally, the “targeted pre-ductal saturations” data was added in a position the user can see. These pieces of information provide relevant context which might help the user when applying neonatal resuscitation procedures in their simulation. Figure \ref{figure:simulation_general} shows a sample view of the simulation room from the trainee's perspective.


\subsubsection{Teleporting between rooms}

\begin{figure}[ht]
\centerline{\includegraphics[width=1\textwidth]{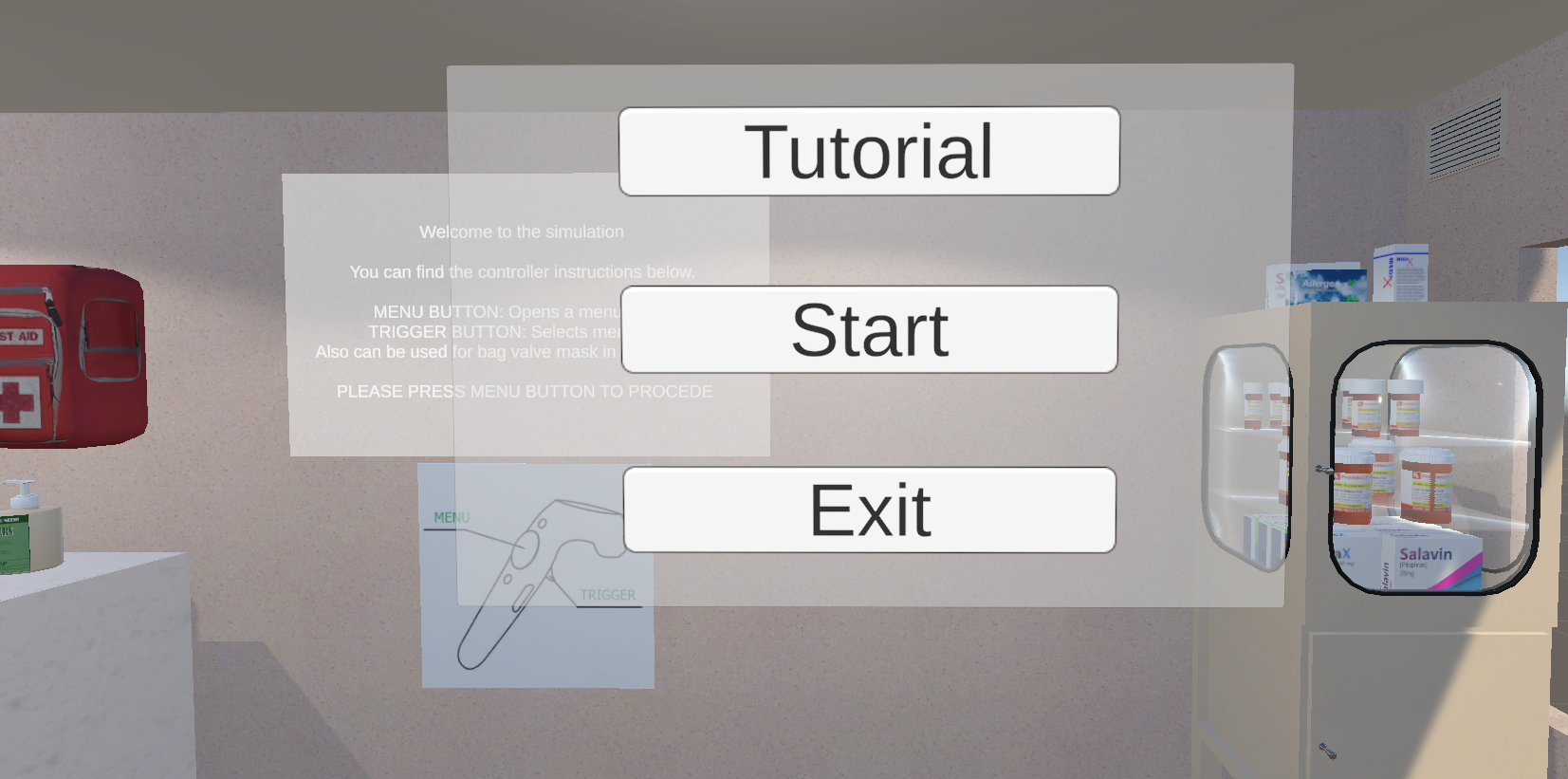}}
\caption{The menu in the simulation. It follows the user's camera and moves wherever they look. Menu items highlight red when the user's controller hovers over them.}
\label{figure:simulation_menu}
\end{figure}

The user can travel between the rooms with the help of a main menu in VR, shown in Figure \ref{figure:simulation_menu}. To open the menu, the user presses the trackpad button of the HTC VIVE Controller in their right hand. After pressing the trackpad, the menu appears in the user's view. There are three options in the menu, which are placed according to the user's point of view; “Tutorial”, “Start” and “Exit”. When the user makes a selection such as "Tutorial" or "Start", it teleports to the corresponding room. 


The teleport feature of the SteamVR package~(\cite{SteamVRUnity}) is used for traveling between rooms. Teleport locations where the user will be located were determined for each room and a unique name tag was created for each of these in Unity. The starting point is pre-defined and fixed in the first room when the user starts the VR simulation. The user can move to different parts of the room by physically moving in real life if they want, but when they teleport to another room, the initial location to go next has been pre-determined. For the tutorial and simulation rooms, this location was determined to be in front of the infant warmer, which is the most suitable position to do PPV.

Apart from the static elements, interactive elements in the simulation included the infant being treated, a supporting doctor's avatar, an infant's warmer, blanket, bag valve mask, pulse monitor, and a virtual version of the user's hand.

\begin{figure}[ht]
\centerline{\includegraphics[width=1\textwidth]{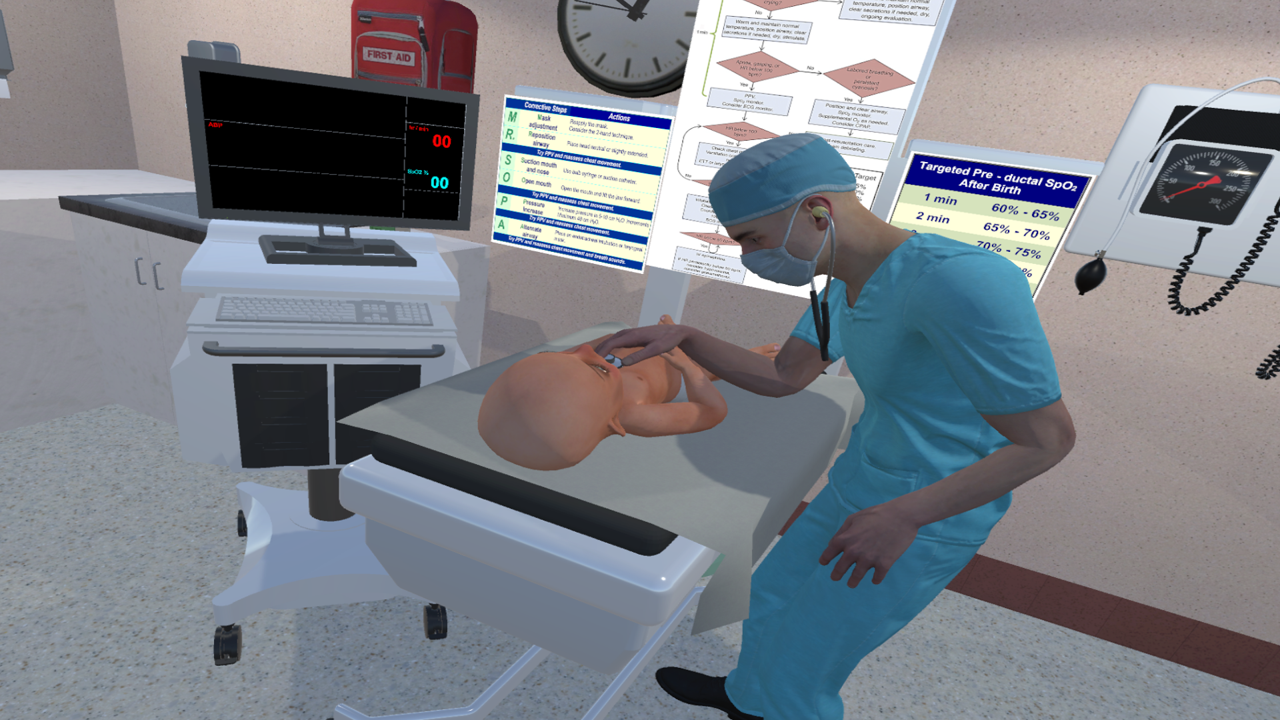}}
\caption{Sample image of the Doctor avatar with stethoscope animation. He stands on the right side of the warmer and helps the NRP provider during the simulation.}
\label{figure:simulation_avatar}
\end{figure}

\paragraph{Doctor Avatar}
Teamwork is very important in the neonatal resuscitation procedure performed on newborns.   As such, in this project, we selected a human avatar to play the role of a doctor teammate to assist the VR user during the training. This human avatar helps the user with animations in important places in the flow of the training and gives voice commands when necessary to prompt the necessary user actions. For animations to be implemented,  a skeletal rig is attached to the joints of this human avatar so that the model moves in a more realistic manner (Figure~\ref{figure:simulation_avatar}).


\paragraph{Infant Warmer}
The infant warmer in which the infant in the simulation will be positioned was requested to be the same model as the infant warmer used at the local NICU. This was done to provide an environment that the participants were familiar with. By the infant warmer, there is a portion of the wall where the NRP procedure posters are hung. These are positioned in front of the user when the user starts the simulation.

\begin{figure}[ht]
\centerline{\includegraphics[width=1\textwidth]{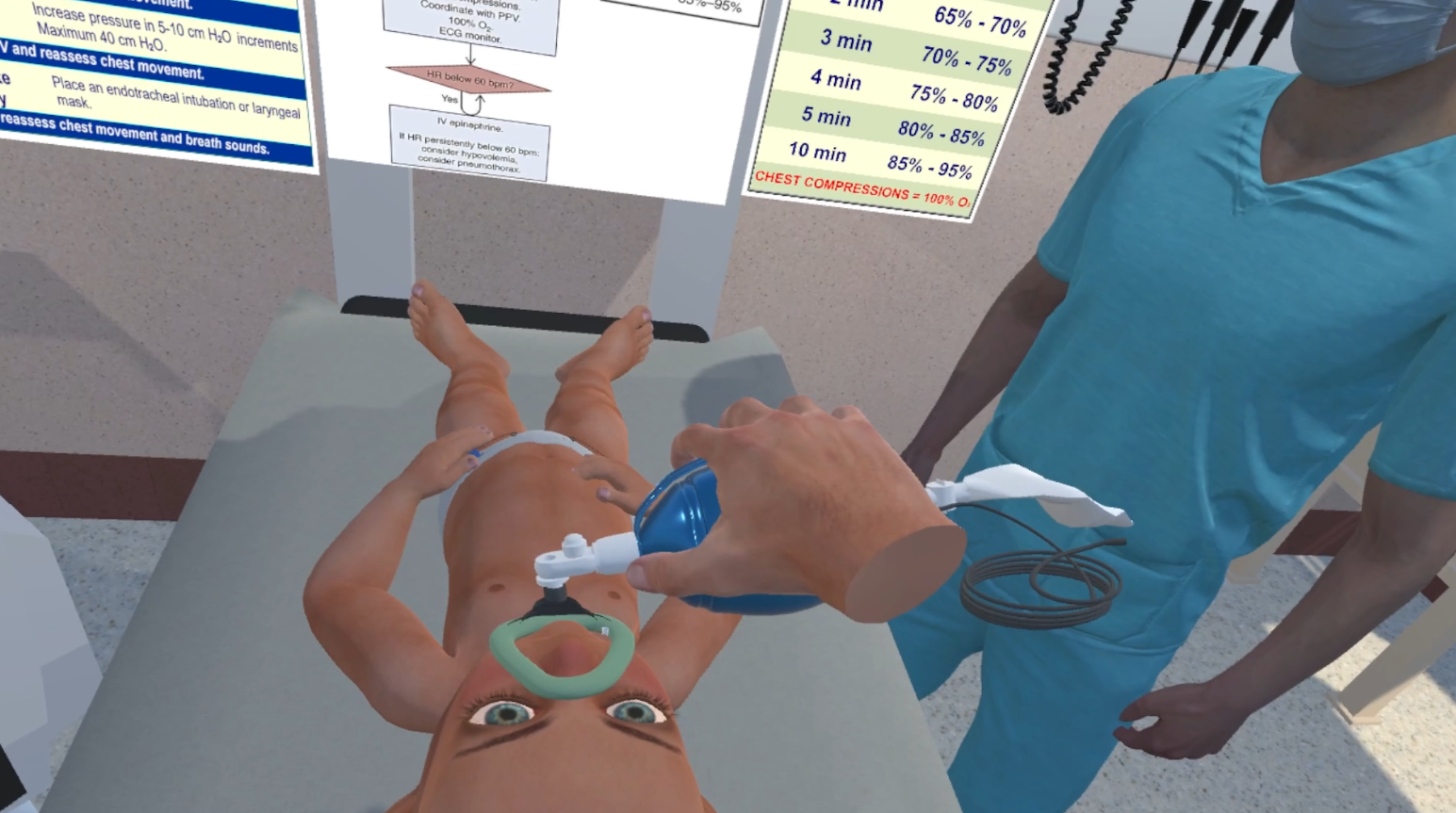}}
\caption{A sample image from the simulation. The user holds the mask to the baby's face and the bag valve mask while doing PPV.}
\label{figure:simulation_resuscitation}
\end{figure}

\paragraph{Bag Valve Mask}
In the event that a baby is not breathing or has a heart rate less than 100 beats per minute, a bag valve mask, one type of positive pressure device recommended by NRP, is used by the user to breath for the baby and provide oxygen as indicated. In the project, the face mask is designed to be held with the user’s left hand, and air/oxygen can be provided by squeezing the bag with the right hand. Therefore, this object consists of two parts; the left part held over the infant's mouth and nose with the left hand and the right part with the right hand to simulate providing air/oxygen via bag valve mask. In the project, a VIVE Tracker was used for the left part, and a VIVE Controller was used for the right part, to hold and squeeze the bag (Figure~\ref{figure:simulation_resuscitation}). When the user squeezes the bag, that is, giving air/oxygen to the infant, the bag is animated through its contraction and expansion. The user does this with the help of the button on the controller while holding the VIVE controller in their right hand. The longer the user holds the button, the more the bag is squeezed.


An HTC VIVE 3D Tracker was used as a bag valve mask replacement held in the left hand of the user and one VR Controller in the right hand to take action on the bag part of the mask. To evoke a realistic sense of engaging in Neonatal Resuscitation, the VIVE Tracker incorporates wrist straps and a ball, strategically positioned in the area intended for the user's grasp of  face mask. This design allowed the user to simulate the physical actions involved in the resuscitation process effectively. 



\begin{figure}[ht]
\centerline{\includegraphics[width=1\textwidth]{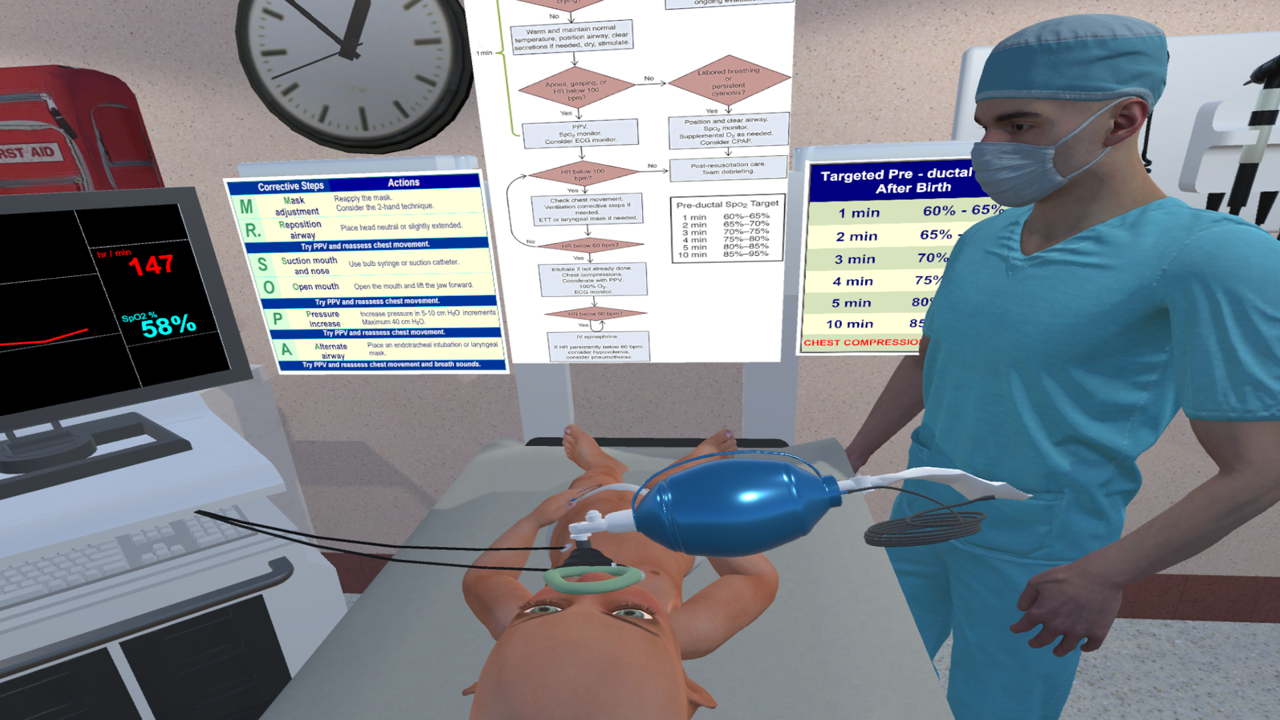}}
\caption{The monitor used to reflect the signals from the ECG sensors attached to the baby’s chest is shown on the left of the image. It is updated with the infant's vitals at every second.}
\label{figure:Simulation_Pulsemonitor}
\end{figure}

\paragraph{Cardiorespiratory  Monitor}
Another interactive object in the simulation is the cardiorespiratory monitor (Figure \ref{figure:Simulation_Pulsemonitor}). The cardiorespiratory monitor is essential in neonatal resuscitation as it provides information on the infant vital signs (heart rate, oxygen saturation). It involves 4 elements; 1) Pulse oximeter, attached to the right wrist of the infant; 2) ECG leads, placed on the infant's chest; 3) ECG cables, designed to connect the leads with the monitor, and lastly; 4) the monitor for the display of the vital signs. 


\paragraph{Trainee's hand}
In order for the user to feel immersed and more belonging in the simulation, a virtual hand (shown in Figure~\ref{figure:simulation_resuscitation}) was used in this project to resemble a real human hand. During the simulation, the user can move this hand as desired in 3D space with the HTC VIVE Controller held in their right hand. The right hand is also used to hold the bag valve mask required for breathing during neonatal resuscitation. The user can squeeze the bag valve mask object that they are holding in their right hand with the help of the trigger button of the VIVE Controller, which prompts a hand squeeze animation.

\section{Validation}

To evaluate the contributions of the VR simulations we performed a randomized controlled crossover study on the effect of VR-NRP on learning outcomes with N=30 NRP trainees. In the study we selected two experimental conditions: 360$^{\circ}$ NRP training videos (a.k.a. 360$^{\circ}$ videos) and the VR-NRP simulation prototype introduced in this article. We chose both of these techniques because they are recent, focus on NRP, and use VR HMDs. 

\subsection{360$^{\circ}$ NRP training videos }

\cite{Curran2021} produced  360$^{\circ}$ NRP training videos and have suggested their use to evaluate the acceptance and benefits of an immersive VR experience in NRP. These videos have received positive evaluations from healthcare providers with experience in NRP and for this reason we used them as the baseline condition to compare against the VR simulation. 

\subsection{Study findings}
In this section we summarize the main findings of the user study. For a  detailed description of the study, its methodology, and experimental results, please refer to the associated article by~\cite{Aydin2024:Study}.

In the study we evaluated the following aspects through questionnaires: 1) Sense of presence, 2) Usefulness, 3) Ease of use, 4) Simulator sickness, and 5) Self-confidence in NRP. We also collected open-ended feedback from the participants. Additionally, participants were evaluated on their performance of the Positive Pressure Ventilation (PPV) resuscitation procedure.

In general, we found that VR technologies are helpful for NRP training of healthcare professionals. VR simulation and 360$^{\circ}$ video were  both perceived as valuable and viewed positively, with overall slightly higher ratings for VR simulation. Participants found both conditions moderately easy to use and reported virtually no simulator sickness effects. 

Participants reported significantly higher perceptions of usefulness (false discovery rate (FDR)-corrected Wilcoxon–Mann–Whitney (WMW) test p=0.014) in particular regarding learning performance and the perceived learning effect of VR. Participants found the VR simulation was more helpful in improving learning performance and enhancing learning. They also reported an increased feeling of presence when experiencing the VR simulation (FDR-corrected WMW test p$<$0.01 for 10 out of 17 questions of the presence questionnaire). Participants using the VR simulation reported higher confidence in certain NRP skills, such as proper mask placement (FDR-corrected WMW test p = 0.0388) and newborn response evaluation (FDR-corrected WMW test p = 0.0173). 

 Overall, participants performed better in providing effective PPV, an essential NRP skill: A blinded assessment of positive pressure ventilation (PPV) skills showed participants exposed to VR performed significantly better in providing effective PPV (FDR-corrected WMW test p = 0.0055). 

The results also suggest that when both techniques are used for NRP training,  360$^{\circ}$ video should be presented first, as a refresher, followed by the VR simulation, to allow for a more hands-on experience. Feedback from participants also suggests that VR-NRP training via VR simulation and 360$^{\circ}$ video could be offered as an adjunct to the in-person training as an additional resource for healthcare providers.

\section{Conclusion and Future Directions}



We have presented VR-NRP, a VR simulation environment designed and developed to refresh the skills of NRP providers. VR-NRP was developed in Unity using the HTC Vive Pro HMD. We compared VR-NRP with 360$^{\circ}$ VR NRP videos, and found that VR-NRP helps improve the learning outcomes of NRP providers, increases confidence in NRP skills, and helps trainees improve their ability to provide effective PPV, a crucial resuscitation skill of NRP.

\subsection{Future Directions}
As suggested by~(\cite{Fealy:2023}), a direction for future research is to explore whether Mixed and Extended Reality could offer additional benefits for NRP training. We believe that by adding tactile feedback to the VR simulation, in particular when practising the bag mask procedures, trainees will develop a stronger sense of realism. The hypothesis is that allowing participants to handle an actual bag and mask that inflates according to the user's actual operation might be more helpful and realistic than the current setup. In open-ended feedback, participants allude to the need for a more hands-on approach to VR training. 

 VR technologies may offer an alternative, cost-efficient modality for increasing access to standardized and portable refresher learning opportunities on NRP concepts. We plan to explore if incorporating both technologies would be helpful in the context of healthcare professionals working in remote or rural areas where in-person training may be difficult to access. A field experiment or evaluation in a rural clinical setting might provide better support for this idea. 

Besides NRP training, VR simulation can be adapted to other training and scenarios, such as cardiopulmonary resuscitation (CPR) training and advanced cardiovascular life support training (ACLS). For our study, we built the VR environment, comprising the room, the characters, the medical equipment and other medical aspects. The scenario and simulation can be easily modified for  CPR and ACLS training in the pediatric and adult patient contexts. After modification of the VR environment, its effectiveness can be assessed for other studies. This research shows that the vast potential of  VR technology and how it can have a beneficial role and impact on healthcare education.


\section{Acknowledgements}
The authors extend their gratitude to Mrs Xiaolin Xu for her support informing the project at its initial stage through a literature review. The authors would also like to thank Claire Bessel and the collaborators at the Office of Professional and Educational Development at Memorial University for their support for the execution of the study and in promoting its results. 

\section{Statements and declarations}

\subsection{Conflicting Interests}
All authors declare they have no conflict of interest. 

\subsection{Funding}
This project was completed with funding from the Janeway Children’s Hospital Foundation (MYA, VC, and OMP) and Memorial University internal grants (MYA, VC, and OMP). The project is also funded in part by the NSERC Discovery Grants Program (OMP).

\subsection{Ethical Considerations}
\label{subsection:ethicsApprovals}
The study was carried out in conformance with Memorial University's policies on Ethics of Research Involving Human Participants. 

\subsubsection{Consent to participate and for publication of findings}
Before the experiment, the participants signed a consent form. In the consent form, the nature of the experiment, its purpose, what it was aimed at, what the participants were asked to do, the questionnaires to be filled, location, length of time, withdrawal from the study information, possible benefits, possible risks, confidentiality, anonymity, data preservation, potential for publication, and communication information was included. 

Participants also received a modest compensation in the form of a gift card. The consent form, recruitment letter, poster, and other related items were approved by the Interdisciplinary Committee on Ethics in Human Research (with ICEHR approval No. 20211585-SC) at Memorial University.

\subsubsection{Guarantor:}
Not applicable.

\section{Author Contributions}
MYA participated in the literature review, designed and implemented the VR simulation, submitted the ethics approval for the project, executed the User Study, evaluated the results, and participated in the production of the manuscript. 
LPC performed the statistical data analysis of the project results and participated in the revision of the manuscript.  
SW assisted in the participant recruitment process, participated in the design of the VR simulation and in the production of the manuscript.
VC was co-supervisor of MYA,  participated in the design and funding of the project, the literature review, and in the production of the manuscript. 
OMP was supervisor of MYA,  participated in the design and funding of the project, the literature review, and the drafting and production of the manuscript.





\printbibliography


\end{document}